# Superconductivity induced by doping Ru in $SrFe_{2-x}Ru_xAs_2$

Yanpeng Qi, Lei Wang, Zhaoshun Gao, Dongliang Wang, Xianping Zhang, Yanwei Ma*

Key Laboratory of Applied Superconductivity, Institute of Electrical Engineering,

Chinese Academy of Sciences, P. O. Box 2703, Beijing 100190, China

**Abstract:**

Using one-step solid state reaction method, we have successfully synthesized the superconductor $SrFe_{1-x}Ru_xAs$. X-ray diffraction indicates that the material has formed the $ThCr_2Si_2$–type structure with a space group I4/mmm. The systematic evolution of the lattice constants demonstrates that the Fe ions are successfully replaced by the Ru. By increasing the doping content of Ru, the spin-density-wave (SDW) transition in the parent compound is suppressed and superconductivity emerges. The maximum superconducting transition temperature is found at 13.5 K with the doping level of x = 0.7. The temperature dependence of DC magnetization confirms superconducting transitions at around 12 K. Our results indicate that similar to non-isoelectronic substitution, isoelectronic substitution contributes to changes in both the carrier concentration and internal pressure, and superconductivity could be induced by isoelectronic substitution.

* Author to whom correspondence should be addressed; E-mail: ywma@mail.iee.ac.cn

## Introduction

Since the discovery of superconductivity in $LaFeAsO_{1-x}F_x$ [1] with $T_c$ about 26 K (denoted as FeAs-1111) last year, the iron based superconductivity has stimulated great interests in the fields of condensed matter physics and material science. Similar to the cuprates, the Fe-As layer is thought to be responsible for superconductivity and La-O layer is carrier reservoir layer to provide electron carrier. By replacing La with other rare earths, $T_c$ can be raised to above 50 K [2-7]. Late on, the oxygen-free iron arsenide compounds $AFe_2As_2$ (denoted as FeAs-122) were discovered and superconductivity was found by appropriate substitution or under pressure [8-13]. Compared to FeAs-1111 type superconductors, the FeAs-122 type superconductors have received special attention due to much simpler structure, without oxygen in the compounds and easy growth of large scale single crystals [14]. It is widely accepted that chemical doping has become a very important strategy to induce superconductivity except under pressure in iron based superconductors. Now there are two mainly categories of doping in the FeAs-122 type superconductor: ⅰ) hole-doping, superconductivity at 38 K by hole doping with partial substitution of potassium for barium [8, 9]. ⅱ) electron-doping, a substitution of Fe ions with Co or Ni can also induce superconductivity with maximum $T_c$ above 20 K [15-17]. It is interesting that superconductivity was realized by doping magnetic element cobalt into the superconducting-active FeAs layers. These results imply that iron based superconductors are tolerant to disorder in the active FeAs layers, unlike the HSTC compounds, where even a small concentration of substitution at the electronically active Cu site is seen to be detrimental to superconductivity [18]. Very recently, superconductivity in Ru substituted $BaFe_{2-x}Ru_xAs_2$ was found [19], furthermore, Ir and Pd substitution Fe sites in $SrFe_2As_2$ compounds have also been exhibited superconductivity [20, 21]. In this paper, we report the synthesis of the new superconductor $SrFe_{1-x}Ru_xAs$ with a $T_c$ of 13.5 K by replacing the Fe with the isoelectronic metal Ru, X-ray diffraction indicates that the material has formed the $ThCr_2Si_2$ –type structure with a space group I4/mmm. The variation of the lattice parameters with Ru substitution demonstrates that the Fe ions are successfully replaced by the Ru. The spin-density-wave (SDW) transition in the parent compound is suppressed with Ru doping, and superconductivity emerges.

## Experimental

The polycrystalline $SrFe_{1-x}Ru_xAs$ samples were synthesized by using a one-step solid state reaction method. The details of fabrication process are described elsewhere [6, 17]. Stoichiometric amounts of the starting elements Sr, Fe, Ru and As were thoroughly grounded and encased into pure Nb tubes. After packing, this tube was subsequently rotary swaged and sealed in a Nb tube. The sealed samples were heated to 850 $^{o}$C and kept at this temperature for 35 hours. Then it was cooled down slowly to room temperature. The high purity argon gas was allowed to flow into the furnace during the heat-treatment process. It is note that the sample preparation process except for annealing was performed in glove box in which high pure argon atmosphere is filled.

The x-ray diffraction measurement was performed at room temperature using an MXP18A-HF-type diffractometer with Cu- $K_{\alpha}$ radiation from 20° to 80° with a step of 0.01°. The analysis of x-ray powder diffraction data was done and the lattice constants were derived. DC magnetization measurement was carried out on a Quantum Design physical property measurement system (PPMS). The zero- field- cooled magnetization was measured by cooling the sample at zero field to 2 K, then magnetic field was applied and the data were collected during the warming up process. The field-cooled magnetization data was collected in the warming up process after the sample was cooled down to 2 K at a finite magnetic field. Resistivity measurements were performed by the conventional four-point-probe method.

## Results and discussion

In order to have a comprehensive understanding of the evolution induced by the doping process, we have measured the X-ray diffraction patterns for all samples. The XRD patterns of the $SrFe_{1-x}Ru_xAs$ are shown in figure 1. It is clear that all main peaks of the samples can be indexed to a tetragonal structure with a space group I4/mmm very well. Almost pure phase is achieved for those samples. By fitting the data to the structure calculated with the software X'Pert Plus, we get the lattice constants. In figure 2, we show *a*-axis, *c*-axis lattice parameters and the volume of unit cell for the $SrFe_{1-x}Ru_xAs$ samples. In the parent phase $SrFe_2As_2$, the lattice constants are a = 3.9207 Å and c = 12.3592 Å. By substituting the Ru into the Fe site, the *a*-axis lattice constant expands a bit, while *c*-axis shrinks slightly, indicating a successful chemical substitution. It is believed that smaller *a*-axis and larger *c*-axis lattice constants would mean smaller bond angle of As-Fe-As. There are several reports about the relation between the transition temperature and the bond angle of As-Fe-As, so the Rietveld refinement of the structural data is urgently needed next step. The change of lattice parameters of in $SrFe_{1-x}Ru_xAs$ compounds is similar to the case of doping the Fe with Ir or Pd in $SrFe_2As_2$ and $BaFe_{1-x}Ru_xAs$ [19-21]. It should be noted that the cell volume remains nearly unchanged compared to the parent compound as shown in figure 2c. In particular, the volume of unit cell increase is only 0.5 % for 70 % substitution. This is in contrast to the case of P substitution at the As site in FeAs-122 system, where a shrinkage of both the *a*-axis and *c*-axis lattice parameters results in a shrinkage of the cell volume [22].

Figure 3 shows the temperature dependence of the electrical resistivity for $SrFe_{1-x}Ru_xAs$ samples with x = 0, 0.3, 0.5, 0.7 respectively. It is clear that the parent phase exhibits a clear anomaly (a sharp drop of resistivity) near 210 K, by doping Ru, the resistivity anomaly corresponding to the SDW transition shifted to a lower temperature (at about 170 K for the sample x = 0.3 ). With a further increase in Ru fraction to x = 0.50, the resistivity anomaly disappeared completely and a superconductive-like transition appeared at about 10 K, however, no zero resistivity was observed down to the lowest temperature, probably because of the much broader transition width in this sample. It is noted that the resistivity drop was not converted to a uprising with the Ru doping, which is different from the Co or Pd doped samples [15, 21]. The sample with nominal composition x = 0.5 shows a clear signature of the superconducting transition with an onset of 13.5 K, leading to zero resistance at the low temperatures. The transition width is ~ 5 K. It is remarked that the Ru content for superconductivity is almost the same with P content needed in the (Ba, Eu)$Fe_2As_{1-x}P_x$ compounds[22, 23], however, the transition temperature is significantly lower than substitution P at As site, which is likely due to a shrinkage of both *a*- axis and *c*-axis lattice parameters in (Ba, Eu)$Fe_2As_{1-x}P_x$ compounds.

Figure 4 shows the temperature dependence of the DC magnetization and the resistivity behavior in the same temperature for the $SrFe_{1-x}Ru_xAs$ samples. The magnetization measurement was carried out under a magnetic field of 20 Oe. We can see the presence of diamagnetism in the Ru substituted samples, which correlates with the occurrence of the resistivity drop seen at low temperatures in Fig.3. We compare the diamagnetic signals in the sample x = 0.3, 0.5, 0.7. A clear diamagnetic signal appears around 12 K in the x = 0.7 sample, which corresponds to the middle transition temperature of the resistivity data. The diamagnetism and presence of zero resistance in the sample with x = 0.7, are the proof that Ru substitution in the $SrFe_2As_2$ compounds lead to superconductivity. A very strong Meissner shielding signal was observed in the low temperature regime, which is similar to the case in $BaFe_{1-x}Ru_xAs$ [19].

The discovery of superconductivity in the doped iron based compounds and subsequent improvement in Tc has generated significant interest in uncovering the mechanisms responsible for this novel superconductivity; however, the superconductivity mechanism in these new superconductors remains unclear yet. The parent compounds of the iron based superconductors show first order structural transition from tetragonal to orthorhombic phase, as well as the spin density wave (SDW) transition at somewhat lower temperature. With doping or application of pressure, the two transitions are suppressed and the superconductivity emerges. Most substitutions that have resulted in superconductivity in the $AFe_2As_2$ systems have been non-isoelectronic implying that the substituent contributes to changes in both the carrier concentration and internal pressure due to difference in ionic size. It is interesting that superconductivity appeared in $EuFe_2As_{1-x}P_x$ by iso-electronic substitution [22]. The observation of superconductivity in this system is attributed to increase in chemical pressure, since the substitution of P seems to decrease both the a-axis and c-axis lattice parameters. In the Ru doping $SrFe_2As_2$ compounds, there is an increase in the *a*-axis lattice parameter, whereas the *c*-axis lattice parameter shows a decrease with substitution. Take the compound x = 0.7 for example, *c*-axis lattice parameter decrease is 2.3 %, however, *a*-axis lattice parameters increase only about 1%, it is clear that the shrinkage of *c*-axis lattice parameter is more obvious compared to the expansion of *a*-axis lattice parameter. So the chemical pressures would be produced with the shrinkage of *c*-axis lattice parameters, which could stabilize superconductivity in the iron based system. Furthermore, the shrinkage of *c*-axis suggests the strengthening of interlayer Coulomb attraction, implying the increase of density of negative charge in FeAs layers by the Ru doping. The calculation in $BaFe_{1-x}Ru_xAs$ compounds shows the contribution to density of states from Fe increases with Ru substitution, indicating that the Fe 3d electrons get delocalized with Ru doping [19]. It is suggested that there is an additional charge transfer from Ru to Fe with Ru doing in the $AFe_2As_2$ system, leading to enhanced 3d electron density near $E_F$. Our results suggest that similar to non-isoelectronic substitution, isoelectronic substitution could induce both sufficient carriers and internal chemical pressure, and superconductivity could be observed finally in iso-electronic substitution system.

On the other hand, theoretical calculations indicated that all five 3d orbitals of Fe atoms contribute to the multiple Fermi surfaces, hence to superconductivity, in iron based compounds [24]. Our data further indicate that superconductivity can be easily induced in the $AFe_2As_2$ family by replacing the Fe sites, regardless of the transition metals of 3d or higher d-orbital electrons. More and more iron based superconductors are discovered, which is helpful to shed light on the mechanisms of these new superconductors.

**Conclusions**

In summary, we have synthesized a series of layered $SrFe_{1-x}Ru_xAs$ compounds with the $ThCr_2Si_2$–type structure. The diamagnetism and presence of zero resistance in the sample with x = 0.7, are the proof that Ru substitution in the $SrFe_2As_2$ compounds lead to superconductivity. Our results suggest that similar to non-isoelectronic substitution, isoelectronic substitution could induce both sufficient carriers and internal chemical pressure, and superconductivity could be observed finally in iso-electronic substitution system.

Note added: At completion of this work we became aware of one paper by W. Schnelle (Cond-mat: arXiv, 0903.4668 (2009).), which reported Ru doping induced superconductivity in $SrFe_2As_2$.

## ACKNOWLEDGMENTS

The authors thank Profs. Haihu Wen, Liye Xiao and Liangzhen Lin for their help and useful discussions. This work is partially supported by the National '973' Program (Grant No. 2006CB601004) and Natural Science Foundation of China (Grant Nos: 50777062 and 50802093).

## Captions

Figure 1 XRD patterns of the $SrFe_{1-x}Ru_xAs$ samples. Almost pure phase was achieved for those samples.

Figure 2 Doping dependence of *a*-,*c*-axis lattice constants and volume of unit cell. It is clear that the *a*-axis lattice expands, while *c*-axis lattice shrinks with Ru substitution. The volume of unit cell remains nearly unchanged.

Figure 3 Temperature dependence of resistivity for the $SrFe_{1-x}Ru_xAs$ samples measured in zero field.

Figure 4 Temperature dependence of DC magnetization for the $SrFe_{1-x}Ru_xAs$ samples. The correlated drop in resistivity for the nominal x=0.7 sample is also shown in the figure.

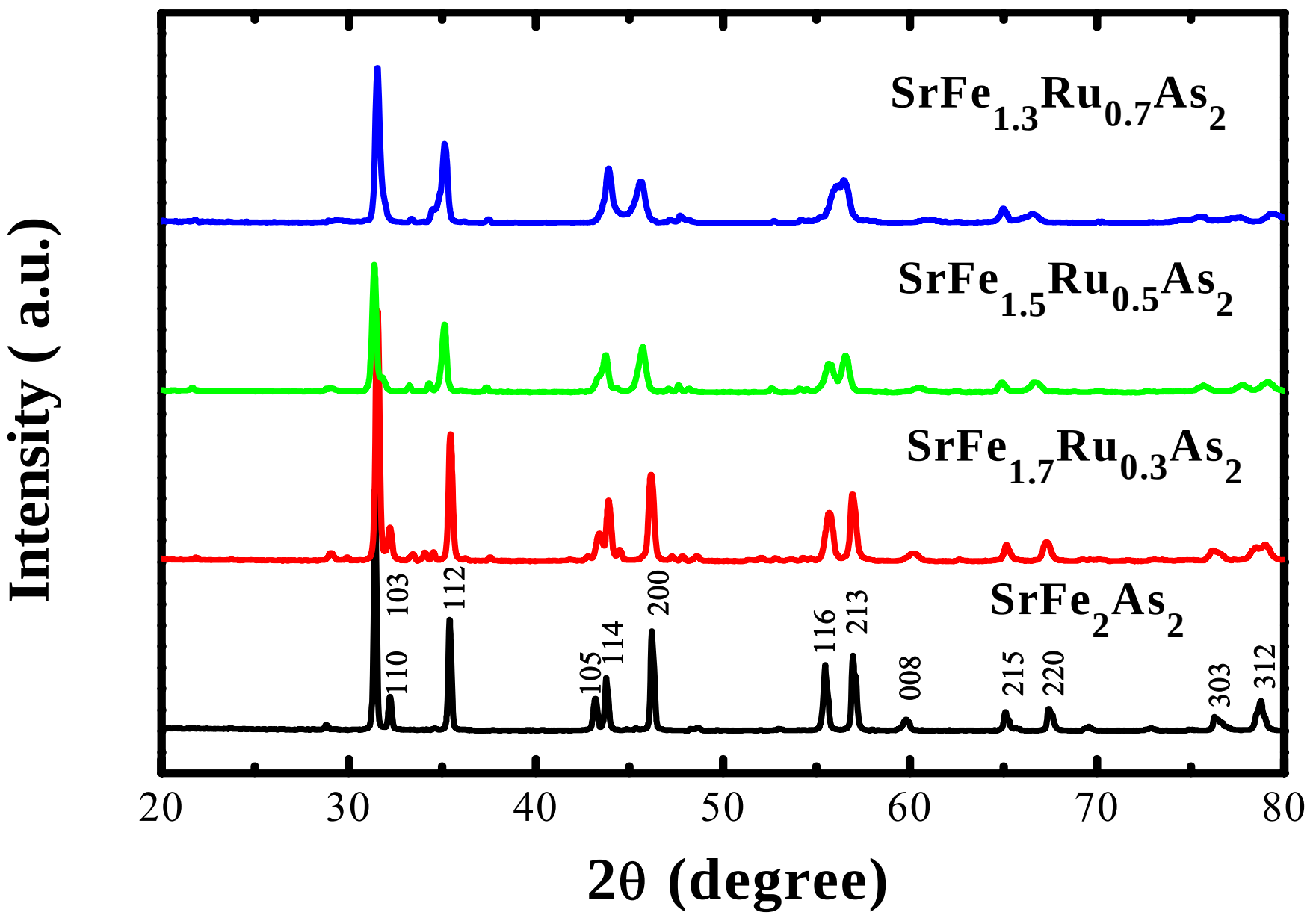

Fig.1 Qi et al.

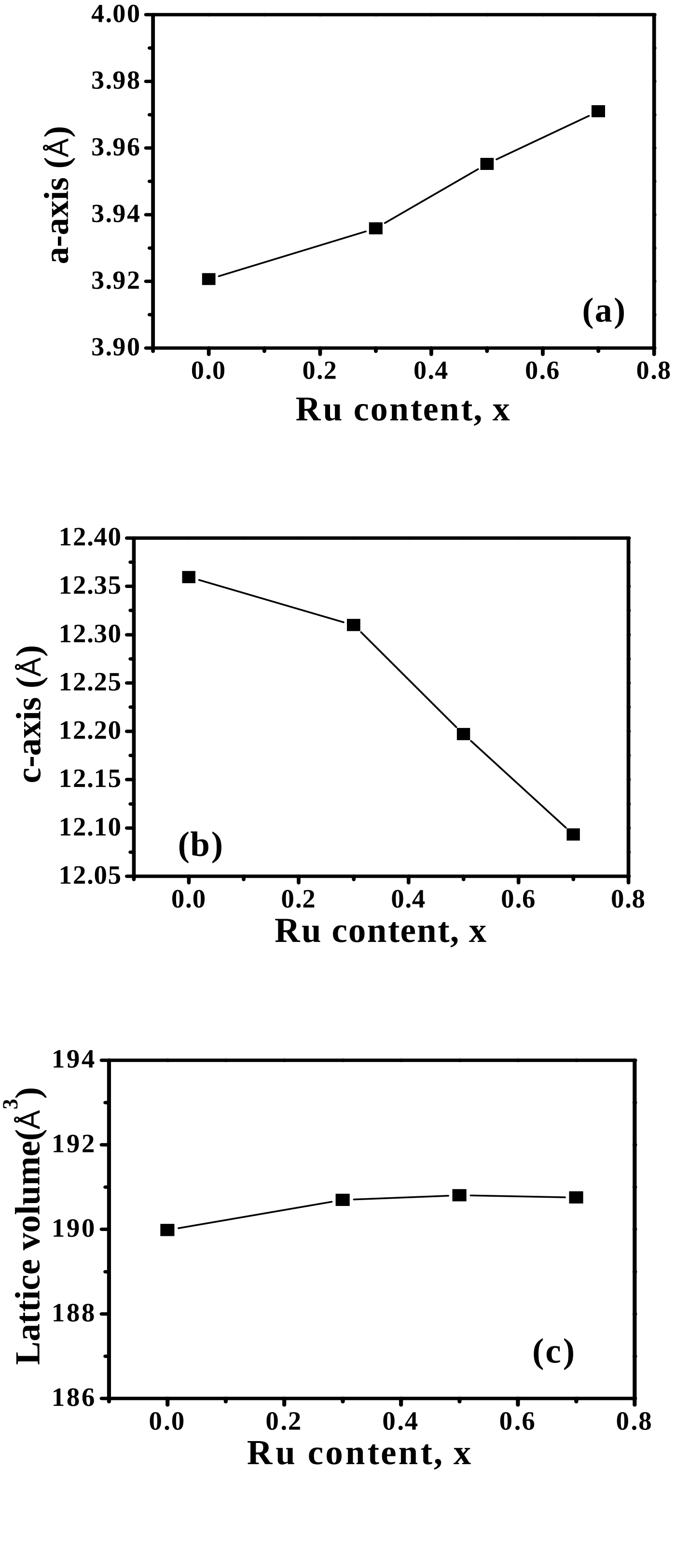

Fig.2 Qi et al.

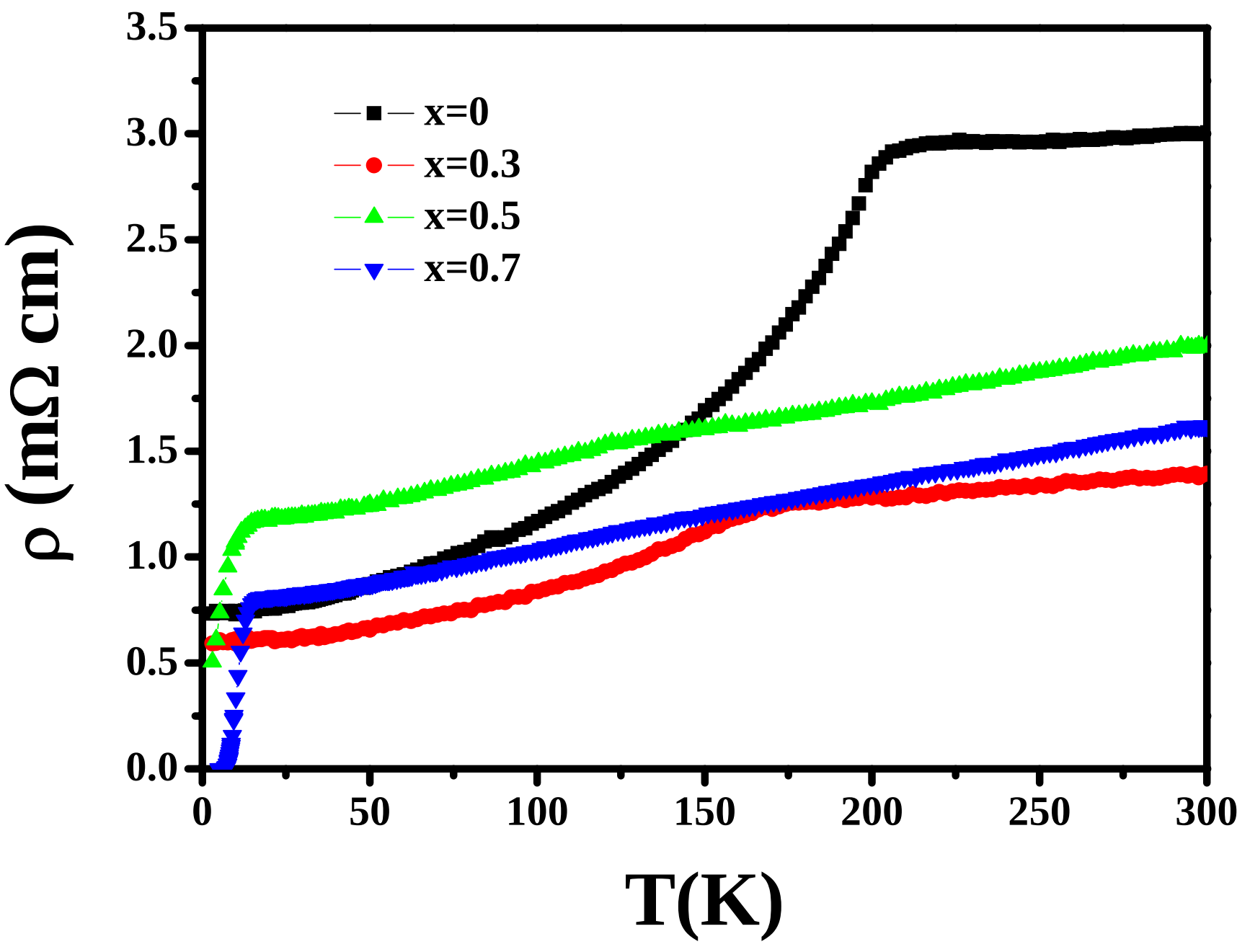

Fig.3 Qi et al.

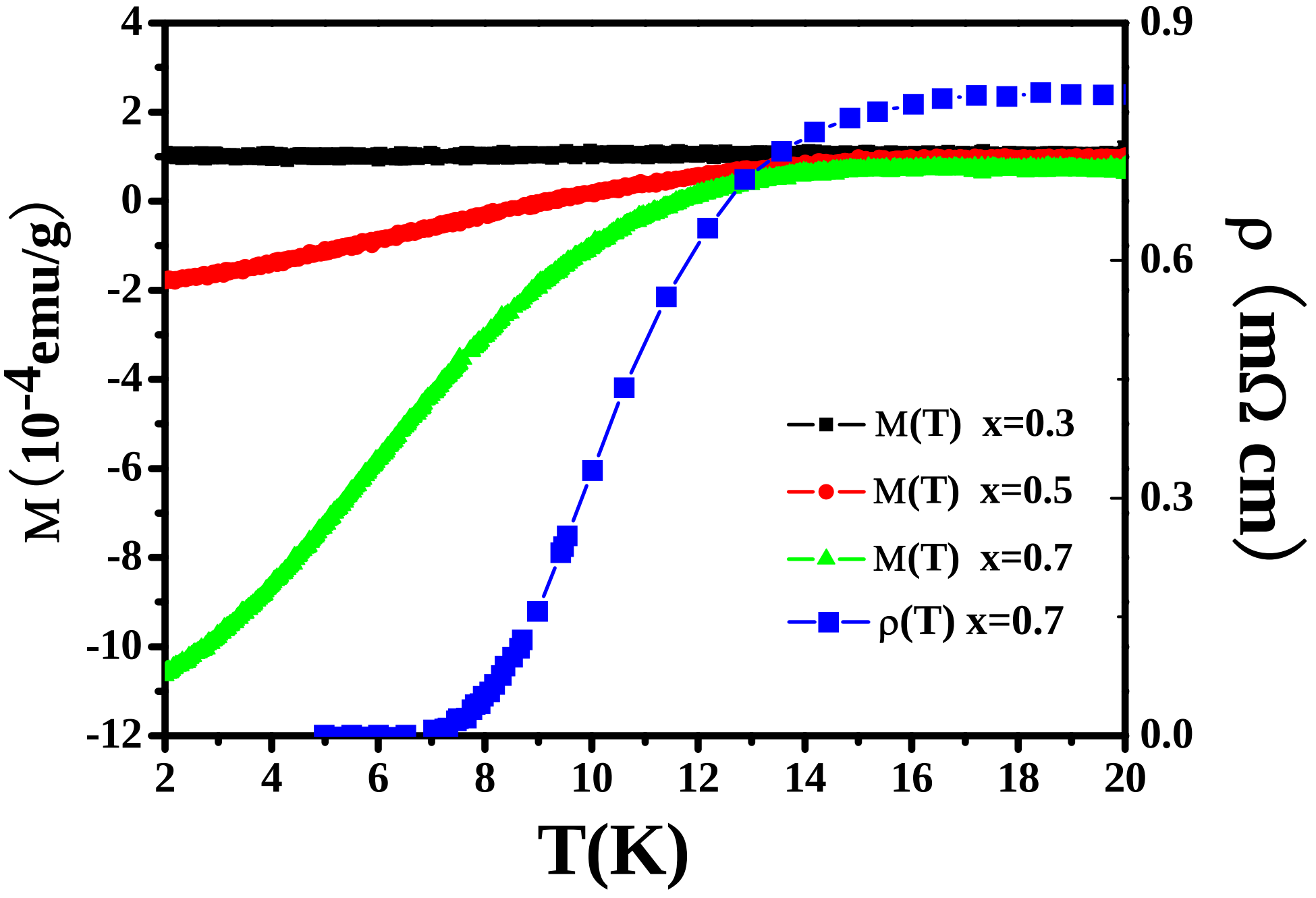

Fig.4 Qi et al.